\documentclass[twocolumn]{revtex4}

\def\al{\alpha}
\def\be{\beta}
\def\ga{\gamma}
\def\ep{\epsilon}
\def\la{\lambda}
\def\om{\omega}
\def\ps{\psi}
\def\si{\sigma}
\def\de{\delta}

\def\fr#1#2{{{#1} \over {#2}}}

\def\bt{\tilde b}
\def\kaf{k_{AF}}
\def\kf{k_{F}}
\def\ka{\kappa}
\def\etal {{\it et al.}}

\newcommand{\beq}{\begin{equation}}
\newcommand{\eeq}{\end{equation}}
\def\prt{\partial}
\def\mn{{\mu\nu}}
\def\half{{\textstyle{1\over 2}}}
\newcommand{\bea}{\begin{eqnarray}}
\newcommand{\eea}{\end{eqnarray}}
\newcommand{\rf}[1]{(\ref{#1})}

\begin{document}

\title{Tests of Lorentz Violation in Atomic and Optical
Physics\footnote{Conference report, at invitation of {\it Comments
on Atomic, Molecular, and Optical Physics (CAMOP)}, on CPT '04, {\it
Third Meeting on CPT and Lorentz Symmetry}, Indiana University,
Bloomington, Indiana, USA, 4-6 August 2004.}}
\author{Neil Russell}
\affiliation{Physics Department, Northern Michigan
University, Marquette, MI 49855, U.S.A.}

\begin{abstract}
Atomic physics can probe the Lorentz and CPT symmetries at
the Planck level.
Bounds on coefficients for Lorentz violation have been found using
atomic clocks,
masers,
electromagnetic cavities,
and Penning traps,
among others,
and in future
it may be possible to place bounds
using spectroscopy of antihydrogen atoms.
The CPT '04 Meeting on CPT and Lorentz Symmetry
was held in August 2004 in Bloomington, Indiana, USA,
and covered Lorentz violation in all branches of physics.
This report gives an overview of the recent advances
in Lorentz-symmetry studies in atomic and optical physics.
\end{abstract}

\maketitle

\subsubsection{Introduction}
Lorentz symmetry is built into the
conventional theories of
particle physics and gravity.
This situation is legitimate,
given the lack of evidence for Lorentz violation.
However, it is possible that nature
is not exactly Lorentz symmetric,
with violations occurring
at a scale too small for past experiments to resolve.
On dimensional grounds,
effects can be expected to involve
the Planck scale,
where gravitational forces are comparable to the
electromagnetic and nuclear forces.
In the last 15 years,
experimental sensitivities have improved rapidly,
and it has become increasingly apparent
that a variety of experiments can access
Planck-scale effects.
In atomic and optical physics,
these include experiments with
atomic clocks,
masers,
optical and microwave resonators,
precision spectroscopy,
and particle traps.
Thus,
Lorentz symmetry or violation
is an experimental question
that needs to be clarified.
This question has received renewed and vigorous attention
stimulated by the introduction of
a framework for quantifying all possible Lorentz violations
called the Standard-Model Extension, or SME \cite{ck,grav}.

The SME is the usual Standard-Model lagrangian
augmented with all possible Lorentz-violating terms
constructed from Standard-Model fields
that are invariant under Lorentz transformations
of the observer's inertial frame.
It allows Lorentz violation
in all areas of physics
from the large-scale gravitational sector
to the small-scale quantum sector.
The theoretical motivation for
considering Lorentz violation
is the possibility that it may occur
in a unified theory of quantum gravity.
At the fundamental level,
one approach is through string theory,
in which Lorentz violation could for example
occur spontaneously \cite{kps}.

Even in its minimal form,
the SME contains several hundred
coefficients for Lorentz violation.
They carry spacetime indices,
and transform as tensors under observer coordinate transformations.
However,
they are fixed entities in spacetime,
and so cannot be controlled in any way.
Distinct SME coefficients
exist for each particle type,
and experiments are sensitive to
differing combinations of coefficients.
Since laboratories are not inertial,
but rotate with the Earth,
one type of Lorentz-violation signal
involves sidereal variations in experimental observables.
The many coefficients
make Lorentz violation possible in a myriad of different ways.
Although any given experiment
can only examine a small part of the coefficient space,
the possibility exists that an experiment from any area
could reveal Lorentz violation.
Dozens of experiments have already probed parts of the coefficient space,
and efforts are continuing to improve precisions.

All the SME coefficients quantify Lorentz violation
and some also quantify CPT violation.
The CPT symmetry, associated with the combined
reversal of charge, inversion of parity,
and reversal of time,
is closely related to Lorentz symmetry
by the CPT theorem \cite{greenberg}.

The Third Meeting on CPT and Lorentz Symmetry
was held in Bloomington, Indiana, in August 2004 \cite{cpt04},
and attracted participants from all parts of the globe.
The conference encompassed experiment and theory
of Lorentz violation from all sectors of physics,
including ones involving
electromagnetic cavities
\cite{photonth,mewesPRD,carroll,kappa,lipa,muller,wolftobar,qbak},
atomic physics
\cite{hbar,maser,hayano,athena,atrap,penning,penning_UW,
penning_Harvard,altschul,akcl,hunter,romalis,HeNeMaser,boosts neutron,
spaceth, muller2},
mesons \cite{hadronexpt},
muons \cite{muexpt},
neutrinos \cite{nu},
and the Higgs \cite{higgs}.
This CAMOP report focuses on the topics specific to
atomic and optical physics.
In particular,
efforts to understand and measure
coefficients for Lorentz violation in the photon and fermion sectors
will be discussed.

\subsubsection{Optical and Microwave Cavities}
On the theoretical front,
the prospect of Lorentz violation
in the photon sector has received much attention \cite{photonth},
and on the experimental side,
various searches for Lorentz violation with electromagnetic cavities
have been performed or are being refined for future experiments.
Typical cavities produce highly stable resonant frequencies
in the optical or microwave regimes.
Stability is monitored by comparison
with a suitable second resonator, often another cavity.
Signals of Lorentz violation include variations
in the output frequency correlated with the orientation
or direction of motion.

The set of possible Lorentz-violating signals
in the electromagnetic sector
is governed by coefficients $(\kf)_{\mu\al\be\ga}$
for CPT symmetry,
and coefficients $(\kaf)^\ka$ for CPT violation \cite{mewesPRD}.
The latter set is bounded at exceptionally
high levels and will be assumed to be zero \cite{carroll}.
This leads to
modified source-free inhomogeneous Maxwell equations:
\beq
\prt_\al{F_\mu}^\al+(\kf)_{\mu\al\be\ga}\prt^\al F^{\be\ga}=0 ,
\label{maxw1}
\eeq
and unchanged homogeneous Maxwell equations:
\beq
\prt_\mu \widetilde F^{\mn}
\equiv \half\ep^{\mu\nu\ka\la}\prt_\mu F_{\ka\la}=0 .
\label{maxw2}
\eeq
Solving these equations for
$F_\mn \equiv \prt_\mu A_\nu -\prt_\nu A_\mu$
with the appropriate boundary conditions for a given cavity
experiment gives the detailed form of
the output frequency dependence
on the $(\kf)_{\mu\al\be\ga}$ coefficients.

Due to the symmetries and other properties
of $(\kf)_{\mu\al\be\ga}$,
there are 19 independent components.
Ten have been bounded by astrophysical data
at impressive levels \cite{kappa},
and the remaining 9 are the focus of cavity experiments
searching for Lorentz violation.
At present,
bounds have been achieved on 7 of these
and sensitivities are steadily improving.

The analysis is aided by defining particular linear
combinations of these components:
five in the traceless symmetric matrix $\tilde\ka_{e-}$,
three in the antisymmetric matrix $\tilde\ka_{o+}$,
and one in the component $\tilde\ka_{\rm tr}$.
These matrices $\tilde\ka_{e-}$ and $\tilde\ka_{o+}$
have spatial indices chosen in any suitable inertial reference frame.
The coefficients for Lorentz violation in the SME
represent geometrical objects fixed in spacetime,
so the frequency output of a cavity oscillator
would undergo cyclic variations as the laboratory
rotates with the motion of the Earth relative
to the fixed distant stars.
In a typical experiment,
the beat frequency of two cylindrical oscillators
mounted at 90 degrees to each other
would have the form
\bea
\fr{\nu_{beat}}\nu &=&
{\cal A}_{\oplus s}\sin\om_\oplus T_\oplus
+{\cal A}_{\oplus c}\cos\om_\oplus T_\oplus
\nonumber \\
&&
+{\cal B}_{\oplus s}\sin2\om_\oplus T_\oplus
+{\cal B}_{\oplus c}\cos2\om_\oplus T_\oplus + {\cal C}_\oplus .
\label{dnu8}
\eea
The components $\tilde\ka_{e-}$, $\tilde\ka_{o+}$,
and $\tilde\ka_{\rm tr}$ appear together with geometrical
factors and an annual time variation in the coefficients
${\cal A}_{\oplus s}, \ {\cal A}_{\oplus c}, \
{\cal B}_{\oplus s}, \ {\cal B}_{\oplus c}$,
and ${\cal C}_{\oplus}.$
An important signal of Lorentz violation
is the time dependence at one or two times the Earth's sidereal
frequency $\om_\oplus$.
To allow comparison of results,
a standard inertial reference frame is used,
involving a Sun-centered
coordinate system $(X,Y,Z)$
with time $T_\oplus$ based on an equinox in the year 2000.

Initial bounds
of about $10^{-13}$ on components of $\tilde\ka_{e-}$
and about $10^{-9}$ on components of $\tilde\ka_{o+}$
were published by a Stanford-based group
in 2003 \cite{lipa}.
These results were achieved with
a pair of cylindrical superconducting cavity-stabilized oscillators
operating in the TM$_{010}$ mode
with one east-west axis and one vertical axis.
These limits have since been improved
by about two orders of magnitude
by one group based at German institutions
in Berlin, D\"usseldorf, and Konstanz \cite{muller},
and by another group
associated with the Paris Observatory
and the University of Western Australia \cite{wolftobar}.

Since one of the signals for Lorentz violation
is a time dependence due to the rotation of the apparatus,
sensitivity can be improved with a rotating turntable
leading to a reduced period of oscillation and other advantages.
All three groups mentioned above are in various stages of
development in this direction.
The Paris-based experiment has compared the output of a cryogenic sapphire
microwave oscillator with a hydrogen maser, both running for several years.
An improved experiment is under way at Western Australia,
involving a rotating turntable with
two sapphire cylinders within superconducting niobium cavities
horizontally mounted with perpendicular axes.
The German group has analyzed more than a year of output from
two orthogonally mounted cryogenic sapphire resonators
running in the optical regime.
A refined experiment includes a precision turntable
and better cryogenics to improve the sensitivity.

Electromagnetostatics
has also been studied \cite{qbak}
for Lorentz-violation signals.
Interesting effects include
a small Lorentz-violating magnetic field
for a stationary point charge,
and a nonzero scalar potential
within a conducting shell containing a magnetostatic source.
Experiments searching for these effects could complement
those done with cavities to study Lorentz violation in the photon sector.

\subsubsection{Testing Lorentz symmetry with fermions}
In the fermion sector,
sensitive tests of Lorentz symmetry
are possible with precision spectroscopy
using masers, atomic clocks, particle traps,
and possibly antihydrogen.
In these systems,
the fixed Lorentz-violating background is quantified for electrons
by the coefficients
$a_\mu^e$, $b_\mu^e$, $H_{\mu \nu}^e$, $c_{\mu \nu}^e$,
and $d_{\mu \nu}^e$,
where the $e$ is for the couplings to electrons
and would be replaced with $p$ or $n$ for protons or neutrons.
The resulting modified Dirac equation
with spinor $\ps$ for an electron of mass $m_e$
and charge $-q$ is:
\bea
( i \ga^\mu D_\mu - m_e &-& a_\mu^e \ga^\mu
- b_\mu^e \ga_5 \ga^\mu
\nonumber \\
- \half H_{\mu \nu}^e \si^{\mu \nu}
&+& i c_{\mu \nu}^e \ga^\mu D^\nu
+ i d_{\mu \nu}^e \ga_5 \ga^\mu D^\nu ) \ps = 0
\, .
\label{dirac}
\eea
In a system such as hydrogen or antihydrogen,
the Coulomb potential is contained in
the vector potential $A_\mu$ and appears in the usual manner,
via $i D_\mu \equiv i \partial_\mu - q A_\mu$.
All the coefficients parameterize Lorentz violation,
and $a_\mu^e$ and $b_\mu^e$
also parameterize CPT violation.
The shifts in the energy levels can be calculated
at leading order in perturbation theory \cite{hbar}.
For example, the shift in the hyperfine c to d transition
of hydrogen in a 0.65-Tesla field is
\beq
 \de\nu^H_{c \rightarrow d}
 \approx - \fr{1}{\pi} (b_3^p - d_{30}^p m_p - H_{12}^p)
 \equiv - \fr{1}{\pi} \, \bt^p_3 \ ,
\label{h_shift}
\eeq
where the superscript indicates that the sensitivity is to proton coefficients.
In this system,
$b_3^p$ is the component of $b_\mu^p$
along the quantization axis
defined by the magnetic-field direction;
similarly,
the 1 and 2 subscripts refer to the other two orthogonal directions
in the laboratory reference frame.

Since the laboratory is rotating with the motion of the Earth,
the appearance of variations in the above hyperfine frequency
with the sidereal period of the Earth
would indicate Lorentz violation.
Experimental bounds on some of the components
$b_X^p$, $b_Y^p$, $b_Z^p$, and $b_T^p$
in the standard inertial reference frame
can be attained by fitting to the appropriate
time dependence.
Sensitivity may be improved with the use
of a rotating turntable.
Sensitivity is also better for transitions with the smallest possible
line width, other factors being equal.
This would indicate that the 1S-2S transition in hydrogen
is a candidate.
However,
calculations show that
this transition is suppressed by a factor
of the square of the fine-structure constant.
Thus, the selection of optimal transitions
involves knowledge of calculated suppressions
and of attainable frequency resolutions.

Recent experiments \cite{maser}
at the Harvard-Smithsonian Center for Astrophysics
have used a hydrogen maser
to place bounds on $\bt_X^p$ and $\bt_Y^p$,
components in the two equatorial directions,
of $2 \times 10^{-27}$ GeV.
The b to d transition was used to search for sidereal variations,
and the result is the sharpest proton-coefficient constraint to date.

When antihydrogen spectroscopy becomes available,
another type of Lorentz test will be possible
using the instantaneous comparison of the hydrogen
spectrum with that of antihydrogen.
In equation \rf{h_shift},
the sign of the coefficient for CPT violation
$b^p_3$ is reversed for antihydrogen.
So, a comparison between corresponding transitions
for the two atoms will isolate the CPT-violating term only.
This clean test cannot be achieved with searches
for sidereal variations,
which bound combinations involving also
coefficients of Lorentz-violating, CPT-preserving terms.
Of the three antihydrogen collaborations at CERN,
the ASACUSA group plans to use an antihydrogen beam to measure
the ground-state hyperfine splitting \cite{hayano}.
The ATHENA \cite{athena} and
ATRAP \cite{atrap} groups
have made progress towards spectroscopy with trapped antiatoms.

Other experiments that compare matter and antimatter
to measure cleanly
coefficients for CPT violation include ones with trapped fermions
in Penning traps \cite{penning}.
The group at the University of Washington in Seattle
placed bounds on SME coefficients based on sidereal
and instantaneous-comparison measurements \cite{penning_UW}.
Other Penning-trap tests have been done and are under development
at Harvard University \cite{penning_Harvard}.
Compton scattering may shed further light
on the behavior of electrons in the Lorentz-violating
background \cite{altschul}.

\subsubsection{Clock-comparison experiments}
The high stability of atomic clocks
is well suited for performing searches for
the sidereal effects of Lorentz violation.
Analysis of cesium and other atoms
common in atomic clocks leads to a number of
challenges not present for simpler systems
such as hydrogen and antihydrogen
because the analysis requires the use of
nuclear models.
However,
an analysis of all possible Lorentz-violation
signals on atomic clocks has been done \cite{akcl}.
This work includes some bounds based
on existing experiments done in other contexts.
For each particle species,
there are five possible coefficient combinations
that these experiments can probe,
one of them being the laboratory-frame component
$\bt^p_3$
in equation (\ref{h_shift}).
This component is one of the four components of
$\bt^p_\mu$,
where $\mu$ refers the laboratory-frame coordinates.
As with all the other experiments mentioned above,
these components have to be related to the
inertial-reference-frame components
$\bt^p_X$,
$\bt^p_Y$,
$\bt^p_Z$,
and $\bt^p_T$
through a transformation that involves
the rotation and speed of the Earth,
the laboratory latitude,
and other geometrical information.
Thus,
the parameter space for Lorentz violation
with clock comparison tests is extensive.
Fortunately,
experiments are sensitive to different
regions of this space,
and rapid progress is being made in probing a large part of it.

Early bounds on SME coefficients
in the fermion sector were obtained using a comparison between
Cs and Hg magnetometers \cite{hunter}
by a group based at Amherst College.

An important aspect of these experiments
is their exceptional frequency stability.
A group at Princeton \cite{romalis}
is developing a $^3$He-K comagnetometer
designed to reduce spin-exchange line broadening,
which should be extremely competitive.

A group at the Harvard-Smithsonian
Center for Astrophysics has developed
a colocated $^{129}$Xe and $^3$He Zeeman maser system.
The one species is used as a magnetometer to
stabilize the 1.5-Gauss magnetic field
while the other species within the same bulb
searches for sidereal variations
due to Lorentz violation.
The resulting bound on the equatorial components
$\bt^n_X$ and $\bt^n_Y$
is at the level of $10^{-31}$GeV \cite{HeNeMaser}.
Recently,
this group has placed the first limits on boost effects
in the neutron sector of the SME \cite{boosts neutron}.
This type of signal indicates Lorentz violation
under boost transformations,
and is suppressed by the ratio
$\be_\oplus = v_\oplus/c = 9.9 \times 10^{-5}$
of the laboratory speed in the standard Sun frame
relative to that of light.

The motion of the laboratory apparatus
relative to the standard Sun-based reference frame
is an important part of almost all Lorentz tests.
It is an advantage to have rotation rates
greater than the sidereal rotation rate of the Earth,
and it is also beneficial to have large velocity changes
relative to the Sun frame.
Space platforms carrying atomic clocks
or other oscillators are therefore interesting
candidates for performing Lorentz tests.
An analysis of such tests based on a satellite orbiting
the Earth has been completed \cite{spaceth}.
The International Space Station
is expected to house various oscillators in the future,
making such tests a possibility.
The microgravity environment
should make it possible to improve on fountain
clocks that are limited by the gravitational
field of the Earth.
Other advantages include
improved access to SME coefficients
in the spatial components perpendicular to the equatorial plane,
and higher rotation rates.
Free-flying missions may offer additional advantages
by optimizing
speeds,
rotation rates,
flight orientation modes,
trajectory geometries,
and other features.

\subsubsection{Discussion}
The SME Lorentz-violation framework encompasses all areas of physics.
This breadth makes it challenging for each subdiscipline
to absorb the progress being made in another,
even though they are not independent.
In fact,
there are many cross-connections between Lorentz violation
from the different sectors of the SME.
For example,
the photons oscillating in a crystal cavity
are affected also by deformations in the oscillator shape
due to the fermions forming the crystal \cite{muller2}.
CPT '04 provided a valuable opportunity
for interaction between
experimentalists and theorists
from diverse sectors including
particle physics,
atomic physics,
gravity,
astrophysics,
and cosmology.
Testing for Planck-scale effects
is a major undertaking,
and no experiment to date
has found evidence of Lorentz violation.
However,
in the last 15 years,
sensitivities of atomic and optical experiments
have improved by many orders of magnitude.
Any experiment that conclusively finds Lorentz violation
will replace the 100-year era of Lorentz symmetry
with a new era in which the minuscule violations
will point towards the fundamental
theory of quantum gravity.

\end{document}